\documentclass{PoS}

\usepackage{here}
\DeclareMathAlphabet{\mathcal}{OMS}{cmsy}{m}{n}
\usepackage{amssymb}
\usepackage{amsmath}
\usepackage{comment}

\title{Probing the dark sector via searches for invisible decays of the Higgs boson at the ILC}

\ShortTitle{Probing the dark sector via searches for invisible decays of the Higgs boson at the ILC}

\author{\speaker{Yu Kato} on behalf of the ILD collaboration\\
        The University of Tokyo, Japan\\
        E-mail: \email{katou@icepp.s.u-tokyo.ac.jp}}

\abstract{
Although the existence of Dark Matter (DM) has been suggested by various astrophysical observations, it has not yet been discovered today.
We can assume a scenario in which the particles that account for the DM can interact with the Standard Model particles only through their couplings with the Higgs sector, the so-called Higgs-portal model.
This model can be investigated by collider experiment using the invisible decay of Higgs boson.
In this study, we evaluate the search ability of International Linear Collider (ILC) for invisible decay of the Higgs using International Large Detector (ILD) full detector simulation.
We estimate 95\% C.L. upper limit (UL) on the branching ratio of invisible Higgs decays and compare them between two center-of-mass energy conditions: $\sqrt{s} = 250$ GeV and 500 GeV.
In addition, we describe the complementarity of lepton collider experiment to the direct detection experiment about DM search ability.

}

\FullConference{
European Physical Society Conference on High Energy Physics - EPS-HEP2019 -\\
			10-17 July, 2019\\
			Ghent, Belgium}

\begin{document}

\section{Introduction}

The Dark Matter (DM) is one of the mysteries left in the Standard Model (SM) of the particle physics.
The existence of DM has been suggested by various astrophysical observations, and many exploratory experiments are currently underway.
Although it has not yet been discovered, there are several constraints from experimental search.

One of the models that describes the interaction between DM and SM particles is the so-called Higgs-Portal model \cite{bib:HiggsPortal}.
In this model, DM in the Universe interact only through their couplings with the Higgs sector.
The simplest and model-independent approach is assuming the minimal Higgs sector, single doublet Higgs field structure, that leads to the SM Higgs boson which has been observed so far and the DM singlets with spin 0, 1 and 1/2.
In this case, the phenomenology of the model would be described only by two parameters in addition to those of the SM: the mass of the DM and its effective coupling to the Higgs boson.

This model can be searched by collider experiments.
In the SM, the branching ratio of the invisible Higgs decay via Z boson decay ($h \to ZZ^* \to 4\nu$) is estimated to be $\sim 0.1\%$.
If there are any new physics beyond the SM, this branching ratio can exceed significantly. 
Note that the mass region of DM that can be searched with this method is less than half of the mass of Higgs boson.



Today, the observed (expected) limit of 19\% (15\%) is set by the CMS group \cite{bib:CMS} and 26\% (17\%) by the ATLAS group \cite{bib:ATLAS} at 95\% C. L. in the LHC experiment.
In addition, the HL-LHC prospect is estimated to be 1.9\% \cite{bib:DM_limit}.
In fact, measurement of missing energy is not easy at hadron collider because the initial state of collision is not clear.
On the other hand, at the lepton collider, such as the International Linear Collider (ILC) \cite{bib:TDR1,bib:TDR2,bib:TDR3_1,bib:TDR3_2,bib:TDR4}, one can search the invisible decay of the Higgs boson with high accuracy using the recoil mass technique because of the clean environment and known initial state.

In this contribution, we evaluate the search ability of the ILC
for invisible decay of the Higgs boson.
ILC is a linear lepton collider which collides electrons and positrons.
Its total length is planned to be 20~km which corresponds to be the center-of-mass energy of $\sqrt{s}=250$ GeV.
The center-of-mass energy is upgradable to 500 GeV and 1 TeV in the future which is a unique advantage of linear accelerators.
In addition, the beam polarization is also powerful tool to suppress backgrounds derived from W boson.

In this study, we use International Large Detector (ILD) \cite{bib:TDR4} which is one of the detector concepts for the ILC.
ILD is designed to optimize Particle Flow Algorithm (PFA) \cite{bib:PFA} which enables to reconstruct and identify all the particles, especially hadron jets.
The jet energy resolution of ILD could be 3 - 4\% using PFA.
We aim to estimate 95\% C.L. upper limit (UL) on branching ratio (BR) of invisible Higgs decay at the ILC with ILD full detector simulation.


\section{Signal and Background}

\begin{figure}[H]
\centering
\includegraphics[width = 6.truecm]{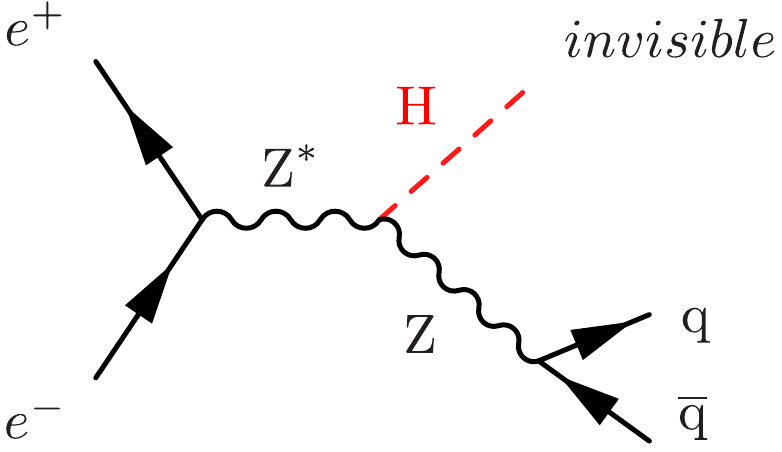}
\caption{The diagram of Higgs-strahlung process.}
\label{fig:zh}
\end{figure}

The signal process of this study, shown in Figure \ref{fig:zh}, is the Higgs-strahlung process with $h \to invisble$ and $Z \to q\overline{q}$. 
This process is expected to be the most sensitive channel because of the high statistics.
Against this signal process, the main background processes are $e^+ e^- \to ZZ \to \nu\overline{\nu} q\overline{q}$ and $e^+ e^- \to WW \to l\nu q\overline{q}$.
The process of leptonic decay of Z boson ($Z \to e^{+}e^{-}$ and $\mu^{+}\mu^{-}$) are also studied and summarized in Ref. \cite{bib:llhinv}.

\section{Simulation Conditions}


We assume two center-of-mass energies, $\sqrt{s}=250$ GeV and 500 GeV, and two beam polarization configurations, $(P_{e^-},P_{e^+})=(\mp 0.8,\pm 0.3)$.
Integrated luminosities of ~900 $\text{fb}^{-1}$ / 1600 $\text{fb}^{-1}$ are assumed for both beam polarization configurations at $\sqrt{s}$ = 250 GeV / 500 GeV based on the ILC running scenario \cite{bib:ILC250}.

We use the signal and background samples which have been generated in the context of the ILC Technical Design Report \cite{bib:TDR1,bib:TDR2,bib:TDR3_1,bib:TDR3_2,bib:TDR4}. 
The beam energy spectrum includes the effects by beamstrahlung and the initial state radiation. 
The beam backgrounds from $\gamma\gamma$ interactions are included in all signal and background processes. 
As the signal process, $e^+ e^- \to q\overline{q}H$ with $h \to ZZ^{*} \to 4\nu$ is used adjusting $BR(h \to invisible) = 10\%$.
The background processes from $e^+ e^-$ interactions are categorized according to the number of final-state fermions and whether Higgs is included: two fermions (2f), four fermions (4f) and SM Higgs processes.

We perform the detector simulation with Mokka \cite{bib:Mokka}/ DD4Hep \cite{bib:DD4Hep}, a Geant4-based \cite{bib:Geant4} full simulation, with the ILD detector model ILD\_o1\_v05 / ILD\_l5\_o1\_v02 for $\sqrt{s}=250$ / 500 GeV.
Events have been reconstructed using \verb|PandoraPFA| \cite{bib:PFA} in the \verb|Marlin| framework \cite{bib:Marlin}.
 

\section{Analysis}

The analysis is performed in three steps: reconstruction, event selection and UL estimation.

\subsection{Event Reconstruction}

The isolated lepton tagging is performed with \verb|IsolatedLeptonTagging| processor \cite{bib:Tagging} to remove events which contain isolated leptons.
We use the parameters summarized in Table \ref{tab:isolep} for \verb|IsolatedLeptonTagging|, where $E_{\mathrm{CAL}}$ is the energy deposit in the whole calorimeter system, $p$ is the track momentum, $E_{\mathrm{ECAL}}$ / $E_{\mathrm{HCAL}}$ is the energy deposit in the electromagnetic / hadron calorimeter system and $E_{\mathrm{yoke}}$ is the energy deposit in the Yoke system.
A multivariate double cone method is used to require isolation and a cut on MVA output is applied.

\begin{table}[H]
\centering
\caption{Parameters for isolated lepton tagging.}
\begin{tabular}{c|cc}
\hline
variable & \multicolumn{2}{c}{condition} \\ 
 & electron & muon \\ \hline
$E_{\mathrm{CAL}}/p$ & 0.5 - 1.3 &$< 0.3$\\
$p$ & $> 5$ GeV & $> 5$ GeV \\
$E_{\mathrm{ECAL}}/(E_{\mathrm{ECAL}}+E_{\mathrm{HCAL}})$ & $> 0.9$ & - \\
$E_{\mathrm{yoke}}$ &-& $> 1.2$ GeV \\
MVA cut for $\sqrt{s}=250$ GeV& $> 0.5$ & $> 0.5$ \\
MVA cut for $\sqrt{s}=500$ GeV& $> 0.8$ & $> 0.8$ \\ \hline
\end{tabular}
\label{tab:isolep}
\end{table}

After the isolated lepton tagging, the jet clustering is applied using Durham algorithm \cite{bib:Durham} by \verb|LCFIPlus| processor \cite{bib:lcfiplus} to force particles into two jets.
At the same time, the beam background rejection with the rejection parameter $\alpha$ of 5.0 is performed at $\sqrt{s} = 500$ GeV to remove the beam backgrounds from $\gamma\gamma$ interactions.

\subsection{Event Selection}

\begin{figure}[h]
	\centering	
    \begin{tabular}{c}

      \begin{minipage}{0.48\hsize}
		\centering		
        \includegraphics[clip, width=8.cm]{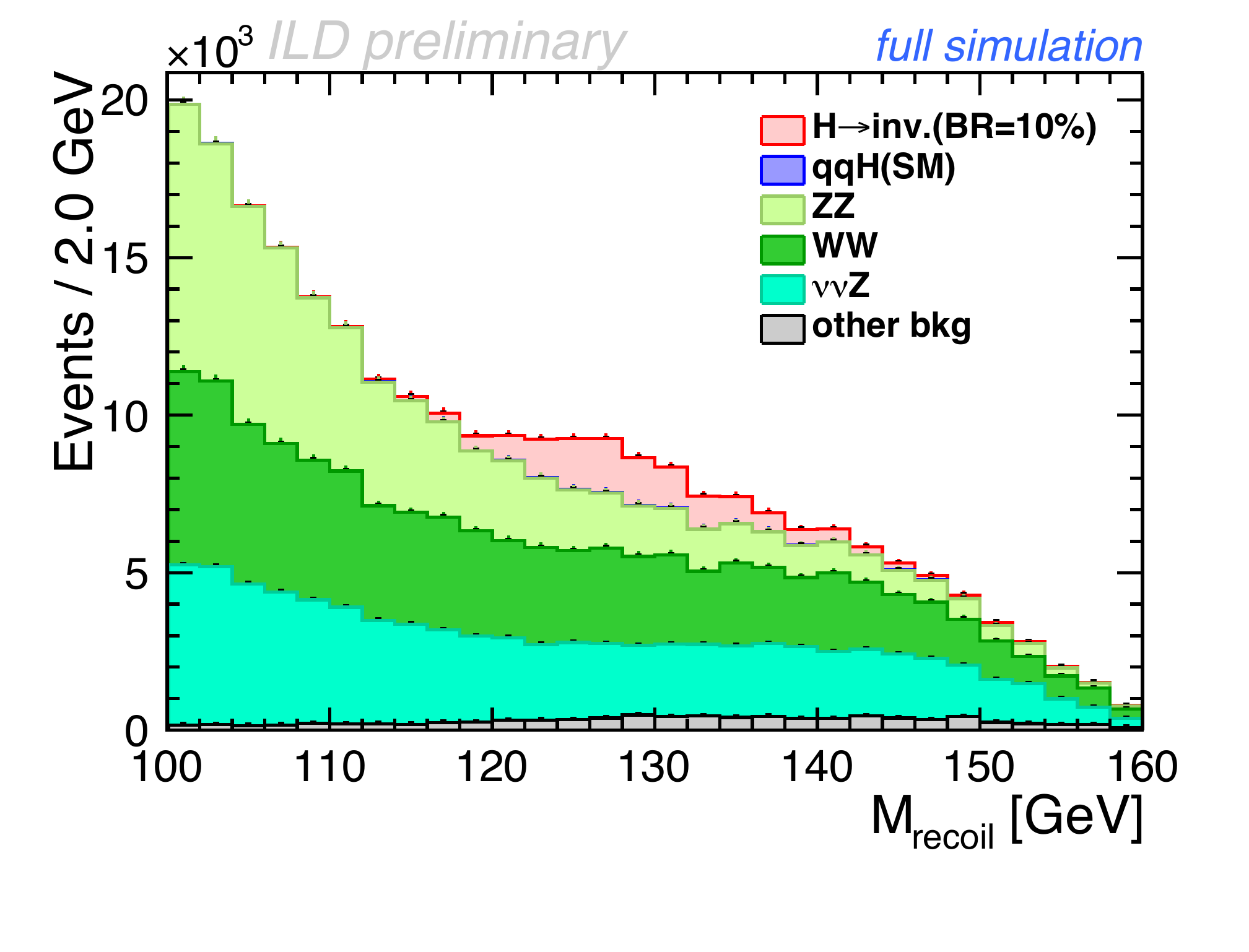}
      \end{minipage}

      \begin{minipage}{0.48\hsize}
		\centering		
        \includegraphics[clip, width=8.cm]{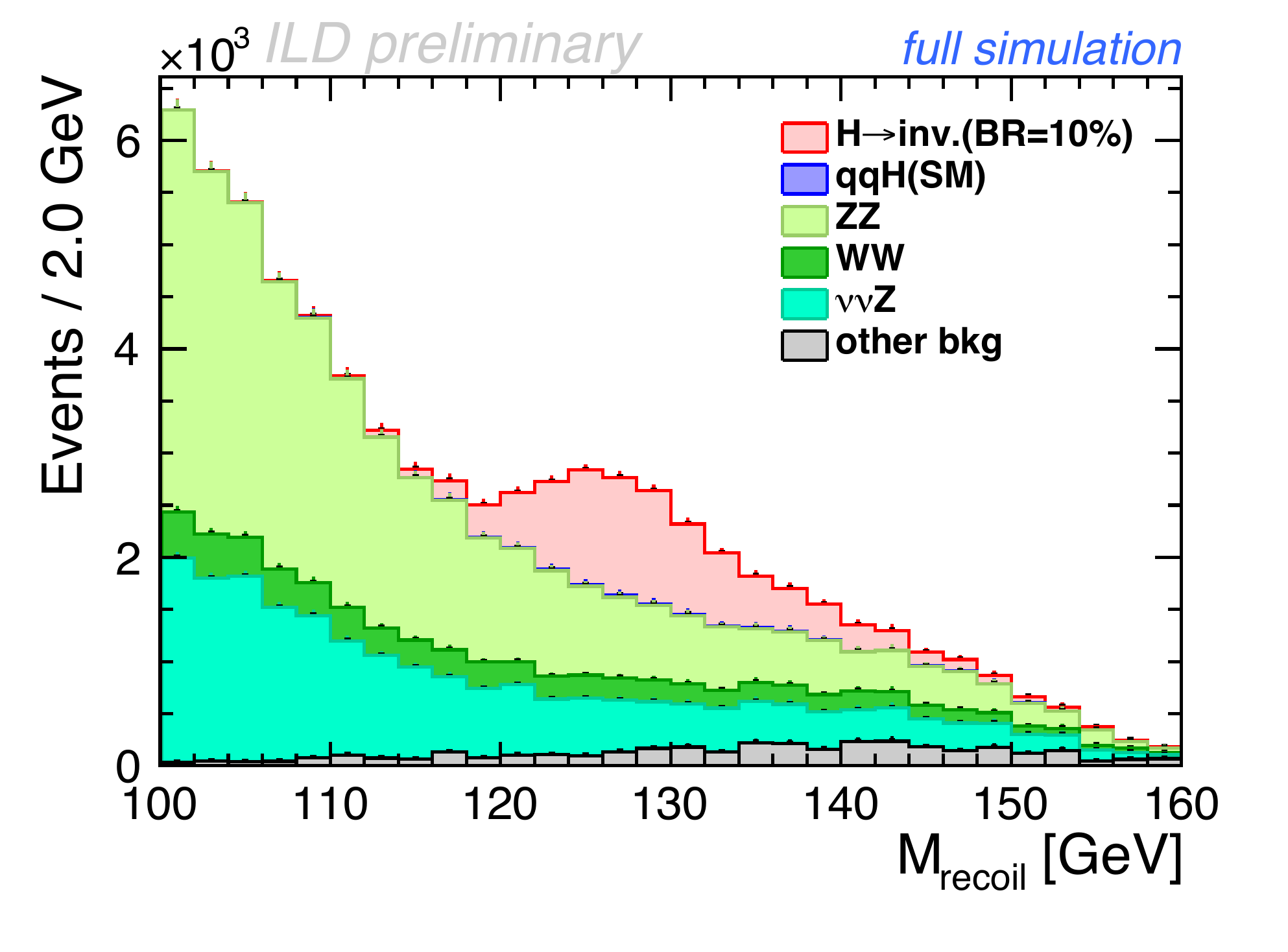}
      \end{minipage}

    \end{tabular}
    \caption{Recoil mass distribution after event selection at $\sqrt{s}=250$ GeV. (left): $(P_{e^-},P_{e^+})=(-0.8,+0.3)$, (right): $(P_{e^-},P_{e^+})=(+0.8,-0.3)$.}
    \label{fig:mrec250}
\end{figure}

\begin{figure}[h]
	\centering	
    \begin{tabular}{c}

      \begin{minipage}{0.48\hsize}
		\centering		
        \includegraphics[clip, width=8.cm]{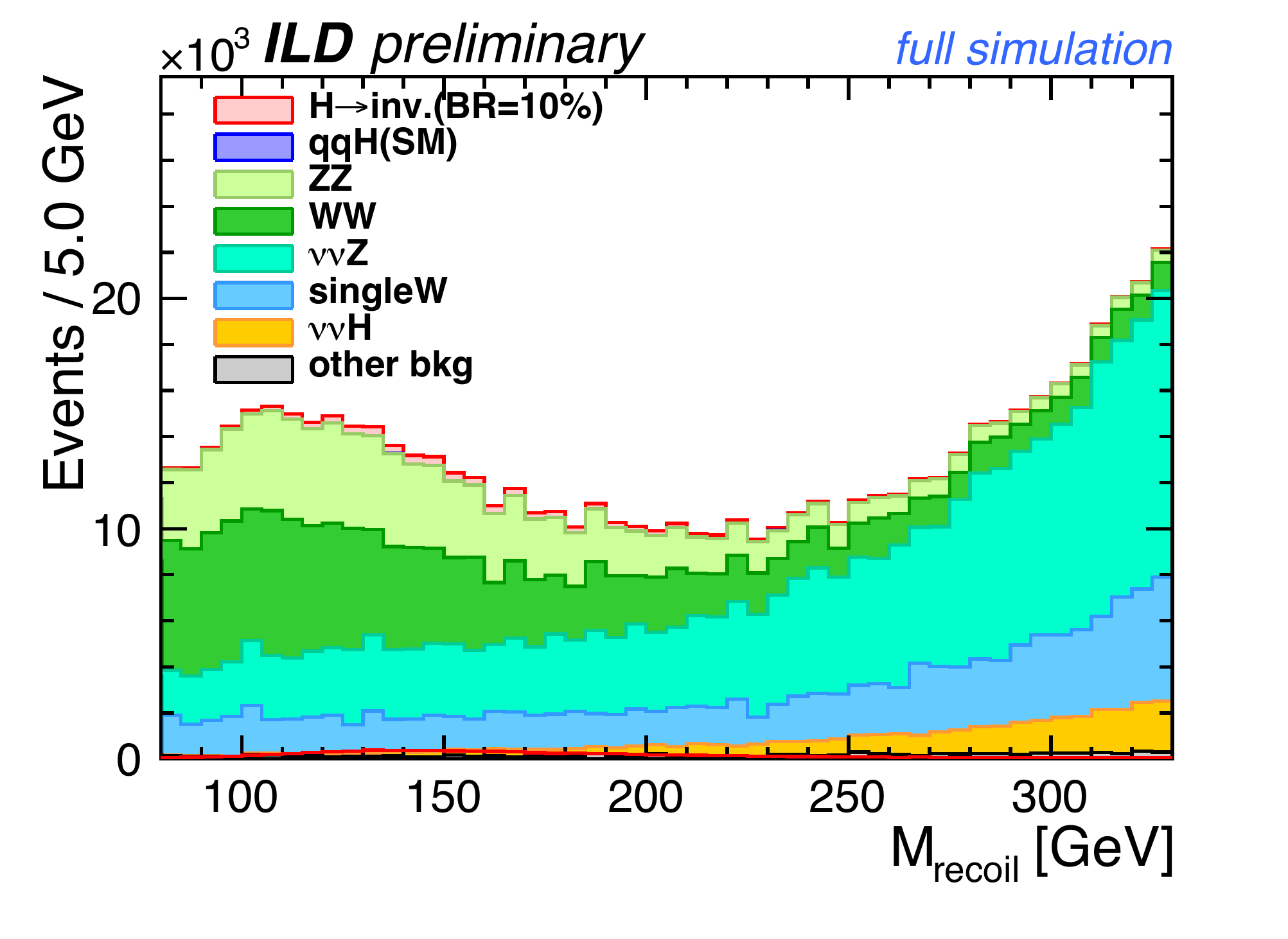}
      \end{minipage}

      \begin{minipage}{0.48\hsize}
		\centering		
        \includegraphics[clip, width=8.cm]{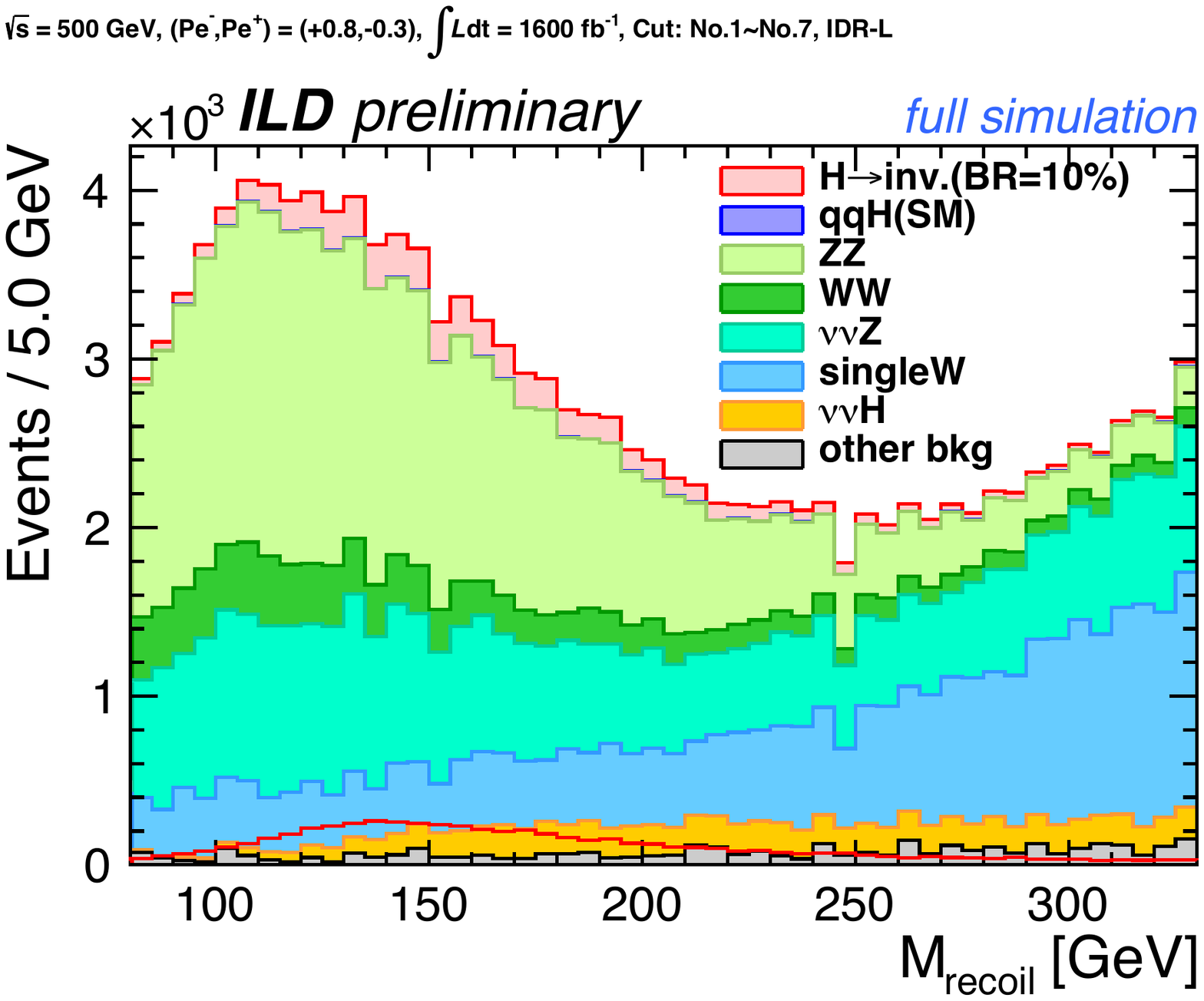}
      \end{minipage}

    \end{tabular}
    \caption{Recoil mass distribution after event selection at $\sqrt{s}=500$ GeV. (left): $(P_{e^-},P_{e^+})=(-0.8,+0.3)$, (right): $(P_{e^-},P_{e^+})=(+0.8,-0.3)$.}

    \label{fig:mrec500}
\end{figure}

After the event reconstruction step, an event selection is performed.
As mentioned above, the branching ratio of $H \to invisible$ is assumed as 10\% for simplicity of event selection.
The cut conditions and number of remaining events are listed in Table \ref{tab:cutl},  \ref{tab:cutr} for $\sqrt{s}=250$ GeV and Table \ref{tab:cutl500}, \ref{tab:cutr500} for $\sqrt{s}=500$ GeV where significance is defined as $N_S/\sqrt{N_S + N_B}$.
And also the recoil mass distribution after event selection is shown in Figure \ref{fig:mrec250} for $\sqrt{s}=250$ GeV and Figure \ref{fig:mrec500} for $\sqrt{s}=500$ GeV.

\begin{table}[H]
	\centering
    \caption{Selection table for $\sqrt{s}=250$ GeV, $(P_{e^-},P_{e^+})=(-0.8,+0.3).$}
    \scalebox{0.85}{
    \begin{tabular}{c||c|c|c}  \hline
		cut condition & signal (efficiency) & all bkg (efficiency) & significance\\ \hline \hline
		No Cut & 18917 (1.000) & 1.417$\times 10^8$ (1.000) & 1.59 \\
		$N_{lep}=0$ & 18880 (0.998) & 9.732$\times 10^7$ (0.687) & 1.91 \\
		Pre-Cut & 18202 (0.962) & 3.358$\times 10^6$ (0.024) & 9.91 \\
		$N_{pfo}>15\&N_{charged}>6$ & 17918 (0.947) & 2.539$\times 10^6$ (0.018) & 11.2 \\
		$p_{Tjj}\in(20,80){\rm GeV}$ & 16983 (0.898) & 1.368$\times 10^6$ (0.010) & 14.4 \\
		$M_{jj}\in(80,100){\rm GeV}$ & 14158 (0.748) & 713194 (0.005) & 16.6 \\
        $|\cos\theta_{jj}|<0.9$ & 13601 (0.719) & 539921 (0.004) & 18.3 \\
       	$M_{recoil}\in(100,160){\rm GeV}$ & 13585 (0.718) & 244051 (0.002) & 26.8\\ \hline
    \end{tabular}}
    \label{tab:cutl}
\end{table}

\begin{table}[H]
	\centering
    \caption{Selection table for $\sqrt{s}=250$ GeV, $(P_{e^-},P_{e^+})=(+0.8,-0.3).$}
    \scalebox{0.85}{
    \begin{tabular}{c||c|c|c} \hline
		cut condition & signal (efficiency) & all bkg (efficiency) & significance\\ \hline \hline
		No Cut & 12776 (1.000) & 7.785$\times 10^7$ (1.000) & 1.45 \\
		$N_{lep}=0$ & 12752 (0.998) & 4.893$\times 10^7$ (0.628) & 1.82 \\
		Pre-Cut & 12270 (0.960) & 1.329$\times 10^6$ (0.017) & 10.6 \\
		$N_{pfo}>15\&N_{charged}>6$ & 12067 (0.945) & 852285 (0.011) & 13.0 \\
		$p_{Tjj}\in(20,80){\rm GeV}$ & 11394 (0.892) & 285847 (0.004) & 20.9 \\
		$M_{jj}\in(80,100){\rm GeV}$ & 9481 (0.742) & 165798 (0.002) & 22.6 \\
        $|\cos\theta_{jj}|<0.9$ & 9126 (0.714) & 130070 (0.002) & 24.5 \\
       	$M_{recoil}\in(100,160){\rm GeV}$ & 9115 (0.713) & 62979 (0.001) & 33.9 \\ \hline
    \end{tabular}}
    \label{tab:cutr}
\end{table}

\begin{table}[H]
	\centering
    \caption{Selection table for $\sqrt{s}=500$ GeV, $(P_{e^-},P_{e^+})=(-0.8,+0.3).$}
    \scalebox{0.85}{
    \begin{tabular}{c||c|c|c} \hline
		cut condition & signal (efficiency) & all bkg (efficiency) & significance\\ \hline \hline
		No Cut & 11147 (1.000) & 9.456$\times 10^7$ (1.000) & 1.15 \\
		$N_{lep}=0$ & 11094 (0.995) & 6.454$\times 10^7$ (0.682) & 1.38 \\
		Pre-Cut & 9910 (0.889) & 4.427$\times 10^6$ (0.047) & 4.71 \\
		$N_{pfo}>15\&N_{charged}>6$ & 9792 (0.878) & 3.959$\times 10^6$ (0.042) & 4.92 \\
		$p_{Tjj}\in(50,250){\rm GeV}$ & 9515 (0.854) & 1.288$\times 10^6$ (0.014) & 8.35 \\
		$M_{jj}\in(80,120){\rm GeV}$ & 8856 (0.794) & 1.050$\times 10^6$ (0.011) & 8.61 \\
       	$M_{recoil}\in(80,330){\rm GeV}$ & 8554 (0.767) & 7.344$\times 10^5$ (0.008) & 9.92 \\ \hline
    \end{tabular}}
    \label{tab:cutl500}
\end{table}

\begin{table}[H]
	\centering
    \caption{Selection table for $\sqrt{s}=500$ GeV, $(P_{e^-},P_{e^+})=(+0.8,-0.3).$}
    \scalebox{0.85}{
    \begin{tabular}{c||c|c|c} \hline
		cut condition & signal (efficiency) & all bkg (efficiency) & significance\\ \hline \hline
		No Cut & 7515 (1.000) & 4.620$\times 10^7$ (1.000) & 1.11 \\
		$N_{lep}=0$ & 7473 (0.994) & 3.096$\times 10^7$ (0.670) & 1.34 \\
		Pre-Cut & 6653 (0.885) & 1.184$\times 10^6$ (0.026) & 6.10 \\
		$N_{pfo}>15\&N_{charged}>6$ & 6580 (0.876) & 982838 (0.021) & 6.62 \\
		$p_{Tjj}\in(50,250){\rm GeV}$ & 6394 (0.851) & 244999 (0.005) & 12.8 \\
		$M_{jj}\in(80,120){\rm GeV}$ & 5953 (0.792) & 196107 (0.004) & 13.2 \\
       	$M_{recoil}\in(80,330){\rm GeV}$ & 5739 (0.764) & 152716 (0.003) & 14.4 \\ \hline
    \end{tabular}}
    \label{tab:cutr500}
\end{table}

\subsection{Upper Limit Estimation}

After the event selection, we estimate 95\% C.L. UL on BR($H \to invisible$).
First, we calculate significance from $N_S$ and $N_B$ for each bin of the recoil mass distribution after selection and combine them in all bins (Root Mean Square).
Then, UL is calculated with combined significance using following formula;

\begin{equation}
    UL_{95\%\:C.L.}(\%) = \frac{10[\%]\times1.65}{significance(BR=10[\%])} \: .
\end{equation}

\section{Results}

The results are summarized in Table \ref{tab:result250} for $\sqrt{s} = 250$ GeV and Table \ref{tab:result500} for $\sqrt{s} = 500$ GeV.
The results of leptonic channel \cite{bib:llhinv} are also summarized in the tables.

\begin{table}[H]
\centering
\begin{tabular}{c}
\begin{minipage}{0.48\hsize}
\caption{95\% C.L. UL on BR($H \to invisible$) for $\sqrt{s}=250$ GeV.}\centering
\scalebox{0.85}{
\begin{tabular}{l|ccc}
\hline
Mode & (-0.8,+0.3) & (+0.8,-0.3) & combined \\
\hline
$Z\to q\overline{q}$ & 0.44\% & 0.31\% & 0.25\% \\
$Z\to l\overline{l}$ & 1.06\% & 0.67\% & 0.57\% \\
combined & - & - & 0.23\% \\
\hline
\end{tabular}}
\label{tab:result250}
\end{minipage}


\begin{minipage}{0.48\hsize}
\centering
\caption{95\% C.L. UL on BR($H \to invisible$) for $\sqrt{s}=500$ GeV.}
\scalebox{0.85}{
\begin{tabular}{l|ccc}
\hline
Mode & (-0.8,+0.3) & (+0.8,-0.3) & combined \\
\hline
$Z\to q\overline{q}$ & 1.30\% & 0.98\% & 0.78\% \\
$Z\to l\overline{l}$ & 2.03\% & 1.48\% & 1.19\% \\
combined & - & - & 0.65\% \\
\hline
\end{tabular}}
\label{tab:result500}
\end{minipage}
\end{tabular}
\end{table}

From these results, one can conclude that the sensitivity is quite better at $\sqrt{s} = 250$ GeV than 500 GeV.
In addition, the hadronic channel ($Z \to q\overline{q}$) has a dominant role for search for invisible Higgs decays.
Moreover, comparing the final result of ILC-250, 0.23\%, and HL-LHC prospect, ~1.9\%, ILC-250 gives a factor of 10 better than HL-LHC prospect.

\begin{figure}[H]
	\centering	
    \begin{tabular}{c}

      \begin{minipage}{0.4\hsize}
		\centering		
        \includegraphics[clip, width=6.4cm]{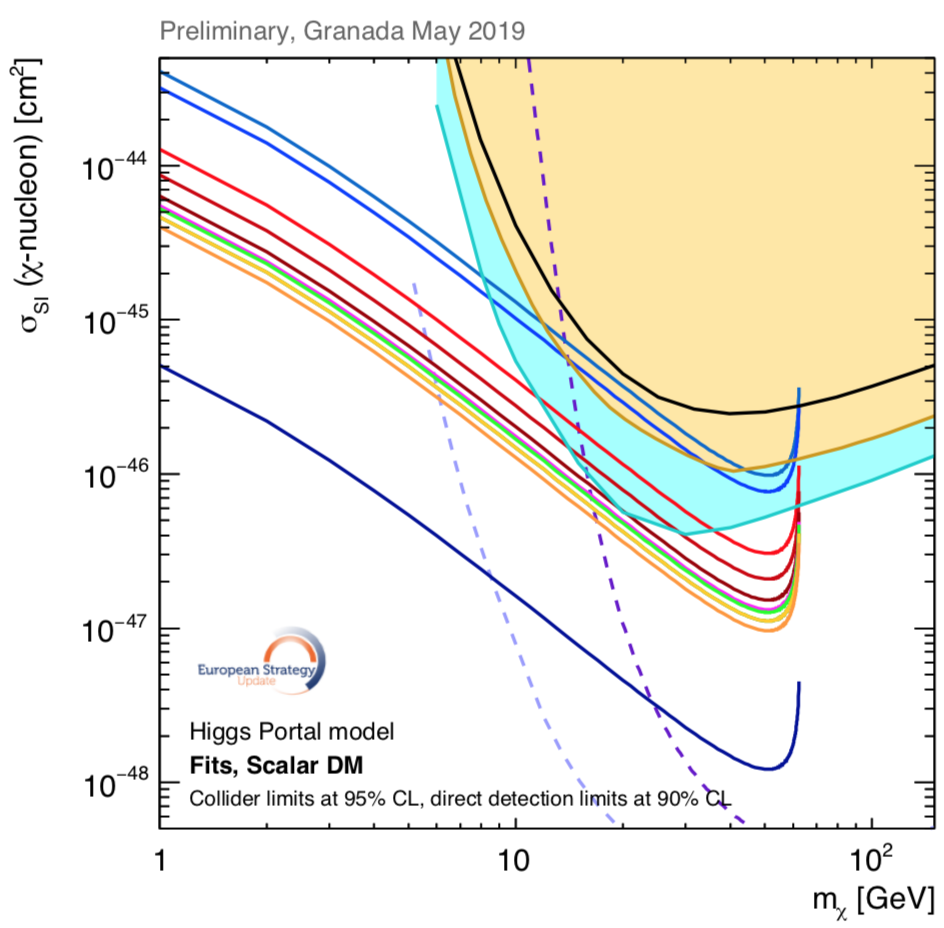}
      \end{minipage}

      \begin{minipage}{0.58\hsize}
		\centering		
        \includegraphics[clip, width=8.0cm]{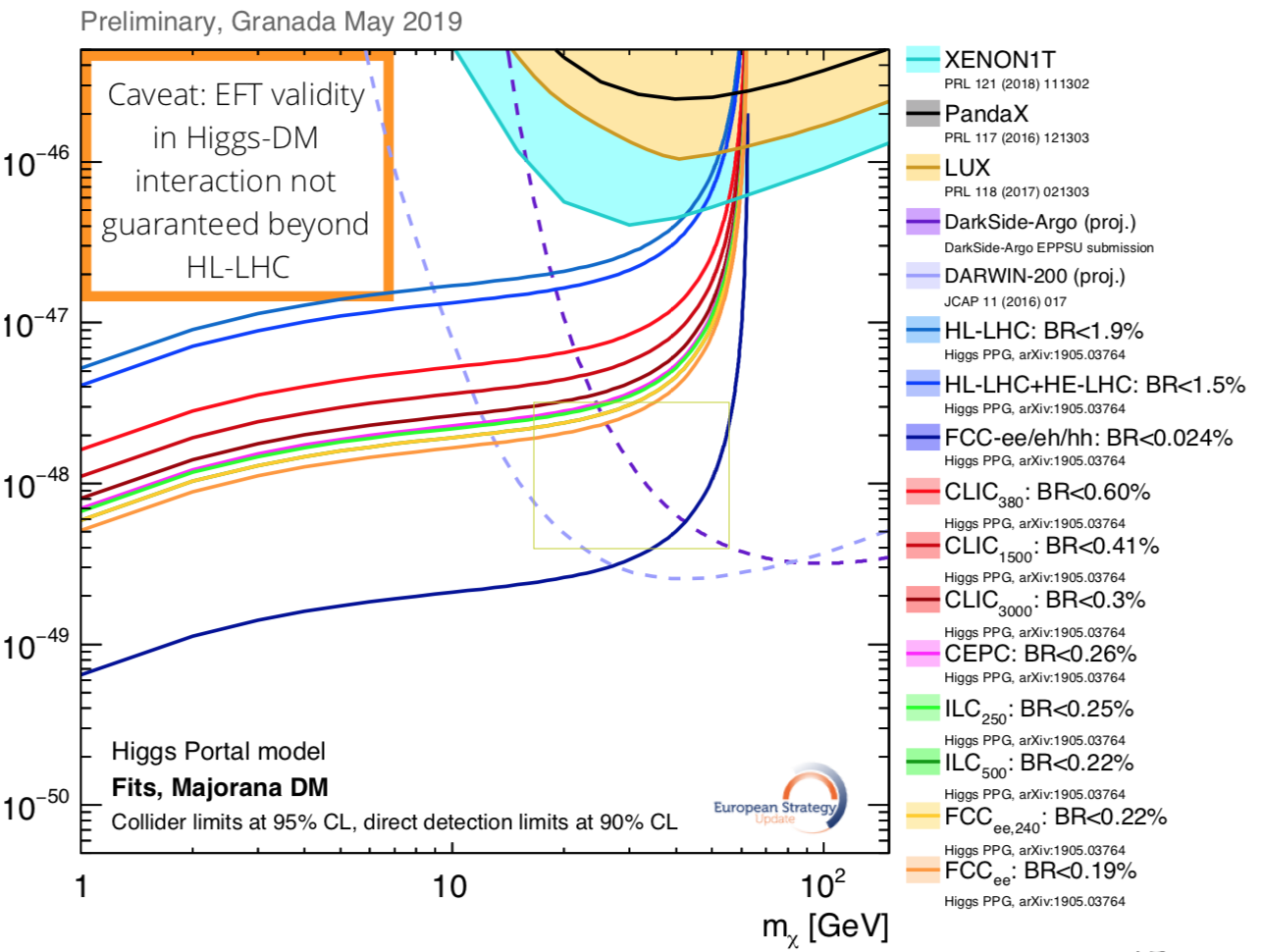}
      \end{minipage}

    \end{tabular}
    \caption{DM limitations. (left): Scalar DM, (right): Majorana DM. Taken from Refs. \cite{bib:ESU, bib:DM_limit}.}
    \label{fig:DM}
\end{figure}

Finally we would like to mention the impact of collider experiments on search for DM.
Figure \ref{fig:DM} shows the limit of cross section between DM and nucleon as a function of DM mass.
The green lines show the ILC results.
One finds that collider experiments are complementary to the direct search experiments. 
The collider experiments have more sensitivity for searching low mass DM than direct search experiments. 
The potential of DM search at the ILC is comparable with other lepton collider projects.

\section{Summary}

We evaluated the search ability of ILC for invisible decay of the Higgs using ILD full detector simulation.
We obtained the 95\% C.L. UL on BR($H \to invisible$) of 0.23\% for 250 GeV ILC.
This limitation is a factor of 10 better than the HL-LHC prospect.
We also compared the DM search possibilities of the other lepton collider projects and direct detection experiments. 
The potential of DM search at the ILC is comparable with other lepton collider projects. The lepton collider projects are complementary to the direct search experiments.


\begin{thebibliography}{99}

\bibitem{bib:HiggsPortal}
G. Arcadi, A. Djouadi, M. Raial, \textit{Dark Matter through the Higgs portal}, \href{https://arxiv.org/abs/1903.03616}{arXiv:1903.03616}.

\bibitem{bib:CMS}
CMS Collaboration, \textit{Search for invisible decays of a Higgs boson produced through vector boson fusion in proton-proton collisions at $\sqrt{s} =$ 13 TeV}, \href{https://arxiv.org/abs/1809.05937}{arXiv:1809.05937}.

\bibitem{bib:ATLAS}
ATLAS Collaboration, \textit{Search for invisible Higgs boson decays in vector boson fusion at $\sqrt{s} =$ 13 TeV with the ATLAS detector}, \href{https://arxiv.org/abs/1809.06682}{arXiv:1809.0668}.

\bibitem{bib:DM_limit}
J. de Blas, M. Cepeda, J. D'Hondt, R. K. Ellis, C. Grojean, B. Heinemann, et al., \textit{Higgs Boson studies at future particle colliders}, \href{https://arxiv.org/abs/1905.03764}{arXiv:1905.03764}.

\bibitem{bib:TDR1}
T. Behnke, J. E. Brau, B. Foster, J. Fuster, M. Harrison, J. M. Paterson et al., \textit{The International Linear Collider Technical Design Report - Volume 1: Executive Summary.} , \href{https://arxiv.org/abs/1306.6327}{arXiv:1306.6327}.

\bibitem{bib:TDR2}
H. Baer, T. Barklow, K. Fujii, Y. Gao, A. Hoang, S. Kanemura et al., \textit{The International Linear Collider Technical Design Report - Volume 2: Physics.} , \href{https://arxiv.org/abs/1306.6352}{arXiv:1306.6352}.

\bibitem{bib:TDR3_1}
C. Adolphsen, M. Barone, B. Barish, K. Buesser, P. Burrows, J. Carwardine et al., \textit{The International Linear Collider Technical Design Report - Volume 3.I: Accelerator \& in the Technical Design Phase.} , \href{https://arxiv.org/abs/1306.6353}{arXiv:1306.6353}.

\bibitem{bib:TDR3_2}
C. Adolphsen, M. Barone, B. Barish, K. Buesser, P. Burrows, J. Carwardine et al., \textit{The International Linear Collider Technical Design Report - Volume 3.II: Accelerator Baseline Design.} , \href{https://arxiv.org/abs/1306.6328}{arXiv:1306.6328}.

\bibitem{bib:TDR4}
H. Abramowicz et al., \textit{The International Linear Collider Technical Design Report - Volume 4: Detectors}, \href{https://arxiv.org/abs/1306.6329}{arXiv:1306.6329}.

\bibitem{bib:PFA}
M. A. Thomson, \textit{Particle flow calorimetry and the PandoraPFA algorithm}, \href{https://doi.org/10.1016/j.nima.2009.09.009}{Nucl. Instrum. Meth. \textbf{A611} (2009) 25}.

\bibitem{bib:llhinv}
J. Tian, \textit{``Sensitivity of Higgs self-coupling in BSM \& Higgs invisible decay using Z->ll''}, Presentation at \href{https://agenda.linearcollider.org/event/6735/contributions/33111/attachments/27216/41412/ZHH_ZHllinv20150411_tianjp.pdf}{the 41th ILC General Meeting of ILC Physics Subgroup}.

\bibitem{bib:ILC250}
K. Fujii, C. Grojean, M. E. Peskin, T. Barklow, Y. Gao, S. Kanemura, et al., \textit{Physics Case for the 250 GeV Stage of the International Linear Collider}, \href{https://arxiv.org/abs/1710.07621}{arXiv:1710.07621}.

\bibitem{bib:Mokka}
P. Mora de Freitas, H. Videau, \textit{``Detector simulation with MOKKA / GEANT4: Present and future''}, LC-TOOL-2003-010.

\bibitem{bib:DD4Hep}
M. Frank et al., \textit{AIDASoft/DD4hep}, webpage: \href{http://dd4hep.cern.ch/dd4hep/}{http://dd4hep.cern.ch/dd4hep/}, 2018.

\bibitem{bib:Geant4}
GEANT4 Collaboration, \textit{Geant4 --- a simulation toolkit}, \href{https://doi.org/10.1016/S0168-9002(03)01368-8}{Nucl. Instrum. Meth. \textbf{A506} (2003) 250}.

\bibitem{bib:Marlin}
F. Gaede, \textit{Marlin and LCCD: Software tools for the ILC}, \href{https://doi.org/10.1016/j.nima.2005.11.138}{Nucl. Instrum. Meth. \textbf{A559} (2006) 177}.

\bibitem{bib:Tagging}
J. Tian, C. Duerig, \textit{isolated lepton finder}, Presentation at \href{https://agenda.linearcollider.org/event/6787/contributions/33415/}{High Level Reconstruction Week}, 2015.

\bibitem{bib:Durham}
S.Moretti, L.Lonnblad, T.Sjostrand, \textit{New and Old Jet Clustering Algorithms for Electron - Positron Events}, JHEP 9808:001 (1998), \href{https://arxiv.org/abs/hep-ph/9804296}{arXive:hep-ph/9804296}.

\bibitem{bib:lcfiplus}
T. Suehara, T. Tanabe, \textit{LCFIPlus: A Framework for Jet Analysis in Linear Collider Studies}, \href{https://doi.org/10.1016/j.nima.2015.11.054}{Nucl. Instrum. Meth. \textbf{A808} (2016) 109}.


\bibitem{bib:ESU}
C. Doglioni, \textit{Dark Matter at colliders}, Presentation at \href{https://indico.cern.ch/event/808335/contributions/3373983/attachments/1843062/3023210/20180513_Doglioni_DM_EPPSU.pdf}{Open Symposium - Update of the European Strategy for Particle Physics}.


\end{thebibliography}
\end{document}